\documentclass[12pt]{article}
\usepackage{pdproc} 

\usepackage{graphicx}

\newcommand{\eV}{\,{\rm eV}}

\newcommand{\MeV}{\,{\rm MeV}}
\newcommand{\GeV}{\,{\rm GeV}}

\newcommand{\m}{\,{\rm m}}

\def\CPTV{{\begin{picture}(18,0)(0,0)\put(0,0){CPT}\put(0,0){\line(3,1){24}}\end{picture}}}

\newcommand{\be}{\begin{equation}}
\newcommand{\ee}{\end{equation}}
\newcommand{\ba}{\begin{array}}
\newcommand{\ea}{\end{array}}
\newcommand{\eq}[1]{~(\ref{eq:#1})}
\newcommand{\NP}{Nucl. Phys.}

\newcommand{\PL}{Phys. Lett.}
\newcommand{\PR}{Phys. Rev.}

\newcommand{\fig}[1]{~{\rm \ref{fig:#1}}}
\newcommand{\tab}[1]{~{\ref{tab:#1}}}

 \renewcommand{\thefootnote}{\fnsymbol{footnote}}

\newcommand{\lascia}[1]{}
\font\tenrsfs=rsfs10 at 10pt
\font\sevenrsfs=rsfs7
\font\fiversfs=rsfs5
\newfam\rsfsfam
\textfont\rsfsfam=\tenrsfs
\scriptfont\rsfsfam=\sevenrsfs
\scriptscriptfont\rsfsfam=\fiversfs
\def\mathscr#1{{\fam\rsfsfam\relax#1}}

\def\Lag{\mathscr{L}}

\def\circa#1{\,\raise.3ex\hbox{$#1$\kern-.75em\lower1ex\hbox{$\sim$}}\,}
\makeatletter
\def\art{\@ifnextchar[{\eart}{\oart}}
\def\eart[#1]#2#3#4#5#6{{\rm #2}, {\em #3 \bf #4} {\rm (#6) #5} ({\em #1})}
\def\hepart[#1]#2{{\rm #2, \em#1}}
\newcommand{\oart}[5]{{\rm #1}, {\em #2 \bf #3} {\rm (#5) #4}}

\begin{document}

\renewcommand{\thefootnote}{\alph{footnote}}
  
\title{NEUTRINO ANOMALIES}
\author{ ALESSANDRO STRUMIA}

\address{Dipartimento di Fisica dell'Universit\`a di Pisa and INFN, Italia\\
{\rm E-mail: Alessandro.Strumia@mail.df.unipi.it}}

\abstract{Solar and atmospheric evidences have been established and can be explained by neutrino masses.
Furthermore, other experiments claim a few unconfirmed neutrino anomalies.
We critically reanalyze the $0\nu2\beta$, LSND and NuTeV anomalies.}
   
\normalsize\baselineskip=16.1pt

\section{Introduction}
Solar and atmospheric neutrino data show that  lepton flavour is violated,
and give the two {\em established} evidences
for physics beyond the SM that we have today
(the SM and its first flaw both appeared in 1968).
Atmospheric, solar, reactor and $\nu$ beam data can be fully explained by
 neutrino oscillations with\cite{venezia}
\begin{equation}\begin{array}{cc}
|\Delta m^2_{23}| = (2.7\pm 0.4) 10^{-3}\eV^2,
&
\sin^22\theta_{23} = 1.00 \pm 0.04,      \\[2mm]
\Delta m^2_{12} = (7.1 \pm 0.6)10^{-5}\eV^2,&
\tan^2\theta_{12} = 0.45 \pm  0.06.
\end{array}\label{eq:oscdata}
\end{equation}
Although the specific dependence on neutrino path-length
and energy predicted by oscillations has not yet been fully tested,
present data exclude all alternative interpretations which have been proposed.


It is plausible that the physics behind present discoveries is a $3\times 3$ Majorana neutrino mass matrix.
This is in fact what one gets adding to the SM Lagrangian higher dimensional operators 
(which parameterize the low-energy effects of new physics too heavy to be directly probed):
$$\Lag = \Lag_{\rm SM} + \frac{m_{ij}}{v^2}\frac{(L_iH)(L_j H)}{2\Lambda_L} + \cdots$$
Maybe we are observing the first manifestation of a new
scale of nature, $\Lambda_L \sim 10^{14}\GeV$,
similarly to what happened in 1896, when operators suppressed by the electroweak scale $v=174\GeV$
were first seen as radioactivity by Becquerel.
LHC will directly explore physics at the electroweak scale in $\sim$2008.

Accessing the neutrino scale could be not so fast.
If the theoretical scheme outlined above is true
we have seen 4 of the 9 Majorana parameters contained in $m_{ij}$.
Planned oscillation experiments seem capable of measuring all 6 Majorana parameters which affect oscillations;
neutrino-less double-beta ($0\nu2\beta$) experiments could discover violation of total lepton number
and measure one more parameter.

Alternatively, neutrino experiments might discover something else which does not fit well in the above scheme.
Present neutrino data show a few unconfirmed anomalies:
\begin{enumerate}
\item A reanalysis\cite{evid} of Heidelberg-Moscow\cite{HM} $0\nu2\beta$ data claims $|m_{ee}|\sim  \eV$.

\item LSND\cite{LSND} claims $\bar{\nu}_\mu \to \bar{\nu}_e$ with small $\theta$
and $\Delta m^2\sim \eV^2$.

\item NuTeV\cite{NuTeV} claims $\nu_\mu /$iron  couplings $\sim 1\%$ away from the SM.

  \item The apparent observation of cosmic rays above the GZK cut-off 
might be related to neutrino masses 
($\nu_{\rm UHE}\nu_{\rm CMB}\to Z$ with
$m_\nu\sim \eV$)
or to problems with energy calibration.
\end{enumerate}
We discuss possible interpretations of the first three hints
(within the SM and beyond) and their signals.
Since this is not established physics, 
we unavoidably touch controversial issues:
I try to present what seems true to me  without hiding problems
and signalling the most controversial points.
Hopefully future data will lead to a definite conclusion, maybe confirming one or more of these anomalies.

\begin{figure}
$$\includegraphics{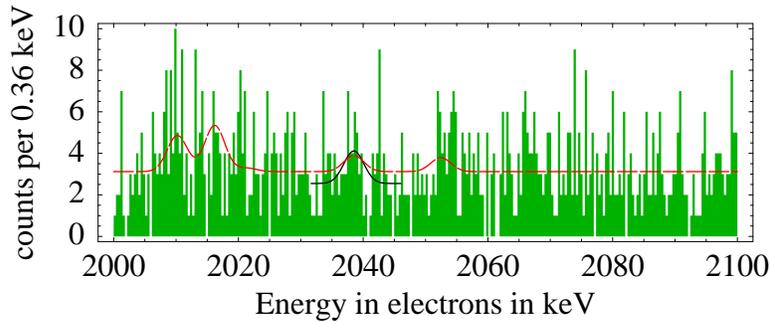}$$
\caption{\label{fig:HM}\em Heidelberg-Moscow $0\nu2\beta$ data
without pulse-shape-analysis.}
\end{figure}

\section{Heidelberg-Moscow}
Neutrino masses distort the end-point spectrum of $\beta$ decays.
From studies of $^3{\rm H}\to {}^3{\rm He}\,e\,\bar\nu_e$
the  {\sc Mainz} experiment 
sets the 95\% CL bound $m_\nu<2.2\eV$  ({\sc Troitsk} has a similar sensitivity)\cite{mainz}.

Some nuclei cannot $\beta$-decay.
For example $^{76}_{32}{\rm Ge}$ cannot $\beta$-decay to
$^{76}_{33}{\rm As}$ that is heavier, so it $\beta\beta$ decays as
$^{76}_{32}{\rm Ge}\to {}^{76}_{34}{\rm Se}\, e \,  e~\bar\nu_e\,  \bar\nu_e$
$(Q=2038.6\,\hbox{keV})$.
Various experiments have observed this and other analogously rare SM processes.
If neutrinos have Majorana masses, the $L$-violating decay
$^{76}_{32}{\rm Ge}\to {}^{76}_{34}{\rm Se}\,e\,e$
has a non zero amplitude proportional to $m_{ee}$, the $ee$ entry of the Majorana neutrino mass matrix.
The experimental signal is two electrons with total kinetic energy equal to the $Q$ value of the decay.
A reanalysis\cite{evid} of the  Heidelberg-Moscow (HM) data\cite{HM} shown in fig.\fig{HM} claimed a
$(2.2\div3.1)\sigma$ evidence for $0\nu2\beta$.
To properly understand HM data one needs to know that
the $0\nu2\beta$ signal is a peak at $Q=2038.6\,\mbox{\rm keV}$ 
with known width,  $\sigma_E \approx 1.6\,\mbox{\rm keV}$ given by the
energy resolution, emerging over
the $\beta\beta$ and other backgrounds, which are not well known.
The evidence in\cite{evid} was claimed assuming
$$\hbox{data} = \hbox{(flat background)} + \hbox{($0\nu2\beta$ peak)}$$
and distinguishing the two components trough a spectral analysis, restricted to data in a small
search window around $Q$ with few $\sigma_E$ size. 
The continuous line in fig.\fig{HM} shows the resulting best-fit:
we obtain a $2.0\sigma$ evidence.
The evidence depends on the choice of the window size:
we have chosen the window where data mostly look like a peak, maximizing
the evidence for peak\footnote{The authors of\cite{evid}
claim that the restriction in the window  search is not a critical arbitrary choice,
in apparent disagreement with\cite{old}.
The claim in\cite{evid} refers to a fit of an average sample of {\em simulated} data (generated under some assumption),
while ref.\cite{old} finds that the restriction in the window search  turns out to be a critical arbitrary choice 
when analyzing the {\em real} data.}.
Employing a large window (which would be the right choice, if the background were really flat)
the evidence decreases to $0.8\sigma$.

\smallskip

Furthermore, the HM spectrum contains a few other apparent peaks, some
around the energies of faint $\gamma$ lines of ${}^{214}\mbox{Bi}$
(a radioactive impurity present in the apparatus).
Fitting all data using all the information we have 
$$\hbox{data} = \hbox{(Bi peaks)}+\hbox{(cte background)} + 
\hbox{($0\nu2\beta$ peak)} $$
we find a $1.2\sigma$ evidence\footnote{The tentative identification of
some of the spurious peaks in the HM spectrum with faint $^{214}$Bi $\gamma$ lines proposed in\cite{evid}
had been criticized in\cite{old,Aalseth}, because
their intensity appeared incompatible with the intense $^{214}$Bi $\gamma$ lines, clearly present in HM data.
This issue has been clarified and the relative intensities are now compatible within $2\sigma$.
The initial estimate has been corrected by
a factor of 6 (as first noticed in footnote 9 of a revised version of\cite{old}, 
there is a trivial normalization error in the published HM data\cite{HM})
times another ${\cal O}(\hbox{few})$ factor related to pile-up effects
(computed assuming that ${}^{214}\mbox{Bi}$ is located in the
copper part of the detector)\cite{Klapdor}.}
as illustrated by the dashed line in fig.\fig{HM}.
Other reanalyses 
performed along similar lines and precisely described by their authors find
$1.1\sigma$\cite{Ianni}, $1.46\sigma$\cite{Klapdor},
less than $1\sigma$\cite{Zde} (these authors combine
HM\cite{HM} and IGEX\cite{IGEX} data.
Both experiments use $^{76}$Ge with a similar energy resolution and background level.
IGEX has about 5 times less statistic and finds a slight 
  deficit of events around the $0\nu2\beta$ $Q$ value, where HM   
  finds a slight excess).
In conclusion, the hint for $0\nu 2\beta$ is 
not statistically significant.
HM data would contain an evidence for $0\nu 2\beta$ if one could show
that the background
around $Q$ is lower than what fig.\fig{HM} seems to indicate.

\begin{figure}[t]
$$\includegraphics[width=7cm]{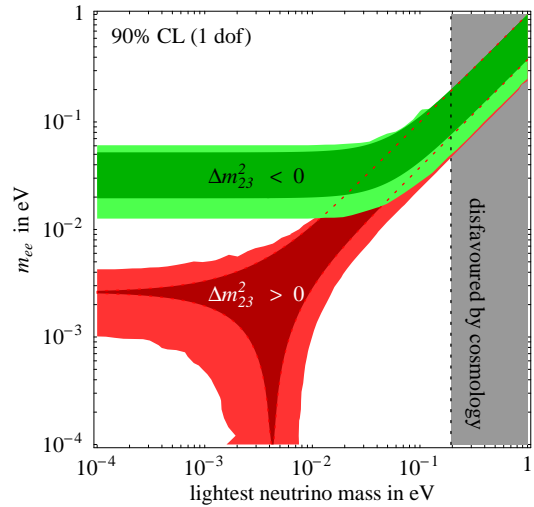}$$
\caption{\label{fig:mee}\em $|m_{ee}|$ range allowed by all oscillation data.}
\end{figure}

\medskip

We now shift topic and study what oscillation data imply on $m_{ee}$,
assuming $3$ neutrinos with Majorana masses.
Therefore we rewrite $|m_{ee}|$ in terms of mixing angles
$\theta_{ij}$, neutrino masses $m_{1,2,3}$ and
Majorana phases $\alpha,\beta$
$$|m_{ee}| = \bigg|\sum_i V_{ei}^2\ m_i \bigg| =
\bigg|\cos^2 \theta_{13}(m_1
\cos^2\theta_{12} +  m_2 e^{i\alpha} \sin^2\theta_{12}) + m_3 e^{i\beta}
\sin^2\theta_{13}\bigg|.$$
Using a global fit of all oscillation data\cite{sunfit},
we plot in fig.\fig{mee} the  $|m_{ee}|$ range at $90\%$ CL
as function of the lightest neutrino mass\cite{old}.
The darker regions in fig.\fig{mee} 
show the remaining uncertainty in $|m_{ee}|$ due to the Majorana phases
that we would achieve if the present best-fit values of
oscillation  parameters in eq.\eq{oscdata} were confirmed with infinite precision
(we are assuming $\theta_{13} = 0$).

Combining the HM\cite{HM} bound on $|m_{ee}|$ 
with solar data which point to less than maximal mixing $\theta_{12}$
one can derive an interesting bound on the mass $m_{\nu}$ of almost-degenerate neutrinos\cite{old}
\begin{equation}
\label{eq:newd}
m_{\nu}< 1.0 ~(1.5)\, h  \eV\qquad\hbox{ at 90 ($99\%$) CL.}
\end{equation}
The factor $h\approx 1$ parameterizes the uncertainty in the 
$0\nu2\beta$ nuclear matrix element (see sect.\ 2.1 of ref.\cite{old}).
Our bound holds under the untested assumption that neutrinos are Majorana particles,
and can be evaded adding e.g.\ Dirac neutrino masses.
Cosmology gives a tighter limit\cite{WMAP,mnu2},
$m_{\nu}  <0.23\ \mbox{eV}$ at $95\%$ CL,
under the untested assumption that
a minimal inflationary model describes structure formation
and can be evaded by  e.g.\ compensating $\nu$ free-streaming with a primordial tilt in the power spectrum.

In the future, the sensitivity of $0\nu2\beta$ experiments to neutrino masses should improve more 
significantly than cosmology and $\beta$-decay.
Fig.\fig{mee} shows that planned $0\nu2\beta$ experiments,
which could reach a sensitivity in $|m_{ee}|$ of few meV,
 should see a signal if the spectrum of neutrinos is
`inverted' (i.e.\ $\Delta m^2_{23}<0$)
or `quasi degenerate' (i.e. $\min m_i \circa{>} 0.05\eV$).

\begin{table}\small
 \renewcommand{\arraystretch}{1.1}
$$\begin{array}{lc|cccc}
\multicolumn{2}{c|}{\hbox{model and number of free parameters}} & \Delta\chi^2& \hbox{mainly  incompatible with} &\hbox{main future test}\\ \hline
\multicolumn{2}{c|}{\hbox{ideal fit (no known model)}}& 0 &&?\\
\Delta L = 2\hbox{ decay }\bar\mu\to\bar e \bar\nu_\mu\bar\nu_e & 6 &  12 &\hbox{\sc Karmen}&\hbox{TWIST}  \\
3+1:~\Delta m^2_{\rm sterile} = \Delta m^2_{\rm LSND} & 9  & 6+9?     & \hbox{{\sc  Bugey} + cosmology?}  &\hbox{\sc MiniBoone}\\
\hbox{3 $\nu$ and \CPTV~ (no $\Delta \bar m^2_{\rm sun}$)}& 10 &  15  & \hbox{KamLAND} & \hbox{\sc KamLAND}\\
\hbox{3 $\nu$ and \CPTV~ (no $\Delta \bar m^2_{\rm atm}$)}& 10 &  25  & \hbox{SK atmospheric} & \hbox{$\bar\nu_\mu$ LBL?}\\
\hbox{normal 3 neutrinos}                             & 5  & 25        &\hbox{LSND}&\hbox{\sc MiniBoone} \\
2+2:~\Delta m^2_{\rm sterile} = \Delta m^2_{\rm sun}  & 9  & 30     & \hbox{SNO}& \hbox{SNO}  \\
2+2:~\Delta m^2_{\rm sterile} = \Delta m^2_{\rm atm}  & 9  & 50    & \hbox{SK atmospheric} & \hbox{$\nu_\mu$ LBL} \\
\end{array}$$ 
  \caption{\label{tab:summary} Interpretations of solar, atmospheric and LSND data, ordered according to
the quality of their global fit.
A $\Delta \chi^2 = n^2$ roughly signals an incompatibility at $n$ standard deviations.}
\end{table}

\section{LSND}
Both the LSND\cite{LSND} and {\sc Karmen}\cite{Karmen} experiments study $\bar\nu_\mu$ obtained from $\mu^+$ decay at rest,
that therefore have a well known energy spectrum up to $52.8\MeV$.
The search for
$\bar\nu_\mu\to\bar\nu_e$ is performed
using the detection reaction $\bar\nu_e p\to n e^+$,
that has a large cross section.
The detectors try to identify both the $e^+$ and the $n$
(via the $2.2\MeV$ $\gamma$ emitted
when the neutron is captured by a proton).
The neutrinos  travel for $L\approx 30\m$ in LSND and
$L\approx 17.5\m$ in {\sc Karmen}.

LSND finds an evidence for $\bar\nu_\mu\to\bar\nu_e$,
that ranges between 3 to $7\sigma$
depending on how  data are analyzed.
This happens because LSND has a poor signal/background ratio:
choosing the selection cuts as in\cite{LSND2}
the LSND sample contains
1000 background events and less than 100 signal events,
distinguished only on a statistical basis.
The statistical significance of the LSND signal depends
on how cuts are chosen,
 and it is crucial that all sources of background have been correctly computed.
The main backgrounds are cosmic rays
and $\nu_e$ misidentification.
The final LSND result
$ P(\bar\nu_\mu\to \bar\nu_e) =   (2.6 \pm 0.8)\, 10^{-3}$
can be explained by oscillations with
$\Delta m^2\circa{>}0.1\eV^2$ 
(see\cite{LSND,LSND2} for the precise range).


 {\sc Karmen} is cleaner than LSND 
but has a few times less statistics
 and shorter base-line.
{\sc Karmen} finds $15$ events versus an expected background of $15.8$ events,
excluding the part of the ($\Delta m^2,\theta$) range suggested by LSND with larger $\Delta m^2\circa{>}2\eV^2$,
while at smaller $\Delta m^2$ the longer path-length makes LSND more sensitive than {\sc Karmen}.
The LSND anomaly is being tested as $\nu_\mu\to \nu_e$ by the  {\sc MiniBoone} experiment,
which has quite different systematics.

\medskip

\begin{figure*}[t]
$$
\includegraphics[width=70mm]{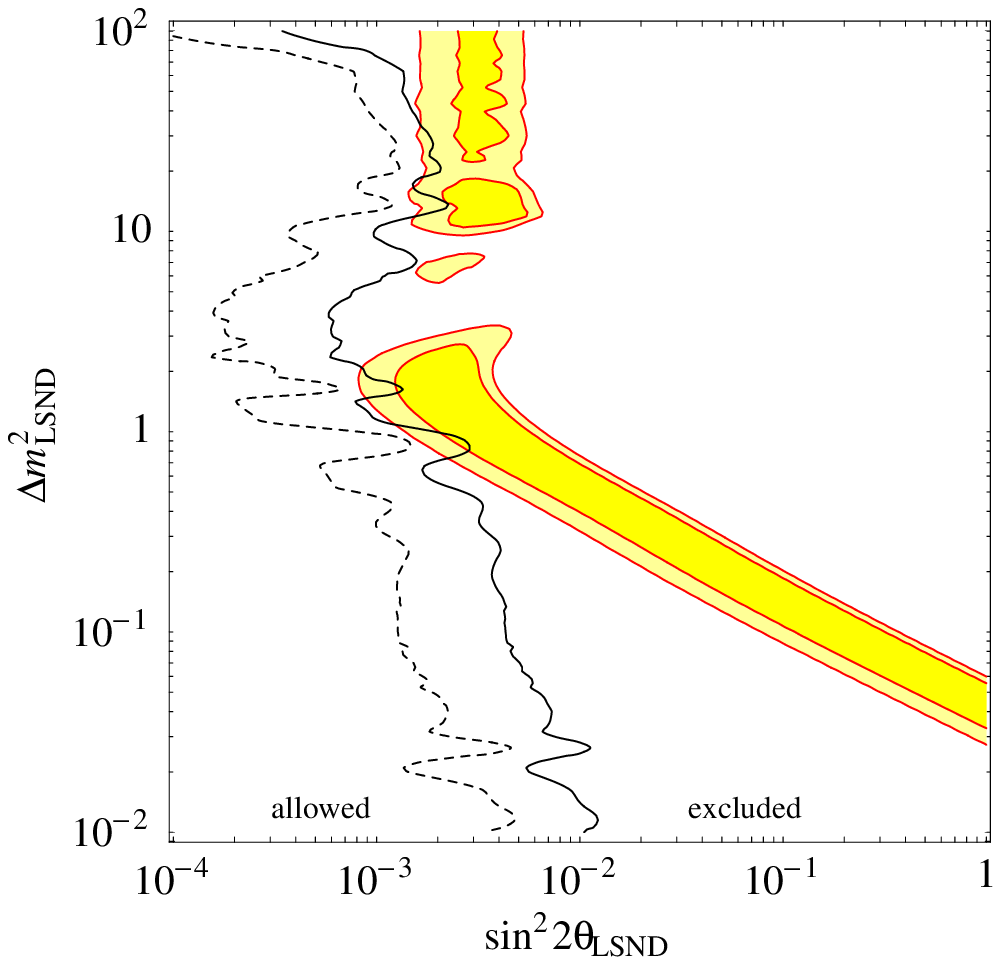} \hspace{1cm}
\includegraphics[width=70mm]{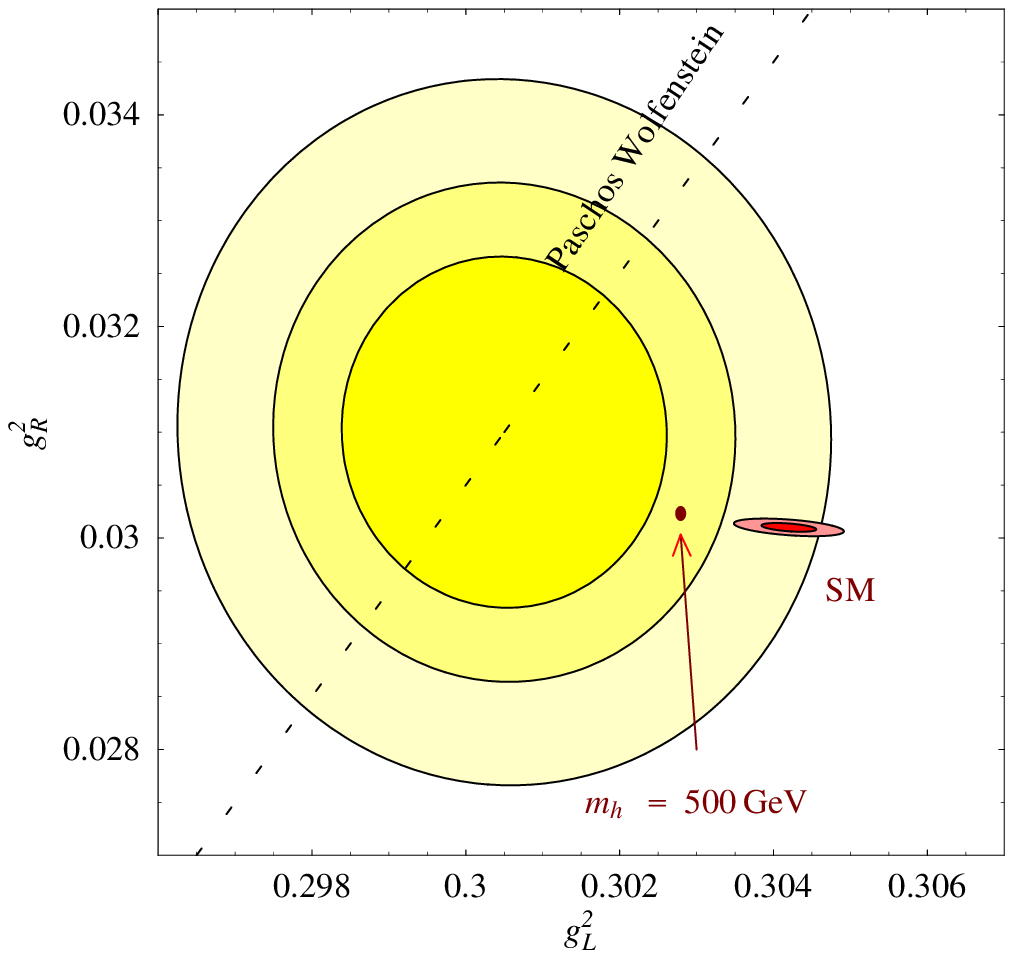}$$
\parbox{7.2cm}{\caption[]{
The LSND region in the ($\sin^22\theta_{\rm LSND},\Delta m^2_{\rm LSND}/\eV^2$) plane
at $90\%$ and $99\%$ CL (shaded area),
compared with the $90\%$ (dashed line) and $99\%$ CL (continuous line)
exclusion bounds from other oscillation data computed 
in a 3+1 oscillation scheme.
\label{fig:split}}}\hfill
\parbox{7.2cm}{\caption[]{\label{fig:NuTeV} The SM prediction for ($g_L^2, g_R^2$) at $68,99\%$ CL compared with the
NuTeV determination at $68,90,99\%$ CL (big ellipses).
The NuTeV central value moves
along the PW line using  different
sets of parton distribution functions that assume $s=\bar{s}$ and $u^p=d^n$.}}
\end{figure*}

If confirmed, the LSND anomaly will require a significant revision of the standard picture.
In fact, oscillations between the three SM neutrinos are described by
two independent squared neutrino mass differences,
allowing to explain only two of the three atmospheric, solar and LSND neutrino anomalies as oscillations.

One possible global explanation of the three anomalies
is that an extra sterile neutrino generates one of them.
The sterile neutrino can be used to generate either the LSND or the solar or the atmospheric anomaly,
or some combination of them.
Only the first possibility is now compatible with data\cite{me,Valle}\footnote{The 
other possibilities would remain strongly disfavoured even if
possible reasons for being more cautious would apply.
\hepart[hep-ph/0210393]{R.\ Foot}
suggests that theoretical errors might have been underestimated in analyses of
NC enriched SK data which disfavour atmospheric sterile oscillations.
Furthermore, bounds on the sterile component involved in solar oscillations would be weakened if solar models underestimate the $^8$B flux.
\hepart[hep-ph/0209373 ]{H.\ Pas et al.} suggest that adjusting all the small mixing angles allowed by
a $4\nu$ framework might weaken the bounds.

\hepart[hep-ph/0210393]{R.\ Foot} suggests that 
scanarios with $\nu_\mu\to\nu_s$ atmospheric oscillations provide a global fit of all neutrino data with
$\chi^2/{\rm dof} \approx 291/276$,
which is acceptable.
This is true, but the goodness-of-fit (gof) test based on the value of the total $\chi^2$ is inefficient when ${\rm dof}\gg 1$:
it may assign an acceptable gof probability to a solution which is already excluded.
This issue was discussed in the context of analyses of solar data in\cite{sunfit}
and can be exemplified by recalling that, according to global fits of solar and KamLAND data\cite{sunfit},
the LOW solution has been excluded but its na\"{\i}ve gof is still acceptable (presently it has 
$\chi^2/{\rm dof} = 89/91$).}.
According to this possibility, named `3+1' oscillations in the jargon, 
$\nu_\mu\to \nu_e$ oscillations at the LSND frequency
proceed trough $\nu_\mu\to \nu_s \to \nu_e$, so that
 the effective $\nu_e\nu_\mu$ mixing angle at the LSND frequency is 
$\theta_{\rm LSND} \approx \theta_{es}\theta_{\mu s}$.
Due to this `product rule'\cite{3+1old} the LSND anomaly somewhat conflicts with
bounds on  $\theta_{es}$ and $\theta_{\mu s}$ from
$\nu_e$ and $\nu_\mu$ disappearance experiments.
The situation is quantitatively summarized in table\tab{summary} and fig.\fig{split}.
The part of the LSND region  still marginally compatible with 
other oscillation experiments has $\Delta m^2_{\rm LSND} \sim \eV^2$
and $\sin^2 2\theta_{\mu s} \sim 0.2$
(for more details see\cite{me,Valle}).

\smallskip

Furthermore, cosmology disfavours such 3+1 interpretation of the LSND anomaly 
 for two different reasons: because
3+1 oscillations thermalize a fourth sterile neutrino,  and because it is heavy.
Since both issues are still controversial, we now try to summarize what  cosmology is really telling.

Massive neutrinos make galaxies less clustered:
recent global fits of cosmological data find $m_\nu \circa{<} 1\eV$
(the analysis performed by the WMAP team gives the strongest bound\cite{WMAP}, 
criticized by\cite{mnu2} who find appropriate a
more conservative treatment of data at small scale and of bias).
The bound on $m_\nu$ gets slightly relaxed if there are $N_\nu = 4$ thermalized neutrinos\cite{mnu2}.
3+1 oscillations indeed thermalize $N_\nu = 4$ neutrinos before the big-bang nucleosynthesis (BBN) epoch,
and $N_\nu = 4$ seems not compatible with BBN.
Assuming standard cosmology, the dominant bound on $N_\nu$ comes from the
BBN prediction for the $^4$He primordial abundancy,
using as input the baryon asymmetry extracted from CMB data.
A global fit of cosmological data 
gives $N_\nu = 2.6\pm 0.2$\cite{mnu2},
which apparently excludes $N_\nu = 4$ and even disfavours $N_\nu = 3$.
One can avoid these conclusions by enlarging the error on the $^4$He primordial abundancy,
since its determination is still controversial.
In table~\ref{tab:summary} we tried to make a quantitative statement, which can
be criticized telling either that it underestimates or that it overestimates the impact of cosmology.

\bigskip

When data started disfavouring interpretations of the LSND anomaly
based on sterile neutrinos,
other more exotic solutions were proposed.
None of them can fully reconcile of data.

Since solar and atmospheric oscillations have been established in neutrinos
but not yet in anti-neutrinos, one can try to explain all anomalies
assuming a CPT-violating neutrino spectrum.
The larger $\Delta m^2$ between $\bar\nu_{e,\mu,\tau}$ is used to explain the LSND anomaly,
and the smaller $\Delta m^2$ could be in the atmospheric  or in the solar range,
at the price of sacrificing some solar\cite{MY} or atmospheric\cite{me,Lyk} anti-neutrino data.
These CPT-violating spectra are now somewhat disfavoured, as
quantitatively summarized in table\tab{summary}.
KamLAND will soon precisely probe CPT in the solar sector.
Doing the same in the atmospheric sector would require
a dedicated  $\bar\nu_\mu$ long-baseline (LBL) experiment
or a dedicated  atmospheric experiment such as {\sc Monolith}.

The LSND anomaly might be produced by a non standard
($\Delta L = 2$!) decay mode $\bar\mu\to\bar\nu_e\bar e \bar\nu$\cite{Babu}.
It affects electroweak precision data and the $\bar e$ spectrum in $\bar\mu$ decays.
The latter signal is under test at TWIST.
This $\bar\mu$ decay interpretation of the LSND anomaly
is disfavoured by {\sc Karmen},
since unlike oscillations it does not exploit the fact that LSND has a longer path-length\cite{me};
a more precise quantitative statement than the one in table~\ref{tab:summary} will be possible if
LSND will analyze their data in this context.

\medskip

In conclusion, the recent experimental progress has disfavoured all proposed interpretations of the LSND anomaly.
Some of them are not yet excluded, and will be tested by future experiments,
as summarized in table~\ref{tab:summary}.

\section{NuTeV}
The NuTeV collaboration\cite{NuTeV} reported a $\sim 3\sigma$ anomaly in
the NC/CC ratio of deep-inelastic muon-neutrino/nucleon scattering.
The effective $\nu_\mu$ coupling to left-handed quarks is found to be about $1\%$
lower than the best fit SM prediction.

The NuTeV collaboration sent both a $\nu_\mu$ and a $\bar\nu_\mu$ beam on an iron target. 
Scattering events were detected by a calorimeter.
The muon produced in CC events gives a long track,
while the hadrons in NC events give a short track.
NuTeV statistically distinguishes NC from CC events putting a cut on the track length.
The ratios of neutral--current (NC) to
charged--current (CC)
deep-inelastic neutrino--nucleon
scattering total cross--sections, $R_\nu$ and $R_{\bar \nu}$,
 are free from the uncertainties on the neutrino fluxes 
and contain the most interesting information.
We recall the tree-level SM prediction for these quantities.
Including only first generation quarks, for an
isoscalar target, and to leading order, $R_\nu$ and $R_{\bar \nu}$ are given by
$$
R_\nu = \frac{\sigma(\nu {\cal N}\to \nu X)}{\sigma(\nu {\cal N}\to \mu X)} = g_L^2 + r g_R^2\qquad
R_{\bar{\nu}} = \frac{\sigma(\bar\nu {\cal N}\to \bar\nu X)}{\sigma(\bar\nu {\cal N}\to \bar\mu X)} = g_L^2 + \frac{1}{r} g_R^2,
$$
where
$$
r \equiv  \frac{\sigma(\bar{\nu}{\cal N}\to \bar\mu
X)}{\sigma({\nu}{\cal N}\to \mu X)}
=\frac{3 \bar{q} +q}{3q+\bar{q}}\sim \frac12$$
and $q=(u+d)/2$ and $\bar q$ denote  the fraction of the nucleon
momentum carried by quarks and antiquarks, respectively.
In this approximation the single parameter $r$ accounts for the uncertain QCD dynamics
and, after including various significant but `trivial' corrections\footnote{Cuts, QED and electroweak corrections, 
iron contains more neutrons than protons, the charm threshold,...
In principle, only a careful job is needed to include all these effects correctly.},
the NuTeV data can be presented as a
measurement of $g_L^2$ and $g_R^2$,
the NC/CC ratio of effective $\nu_\mu q_L$ and $\nu_\mu q_R$ couplings, 
predicted by the SM to be
$$
g_L^2= \frac{1}{2}-\sin^2\theta_{\rm W}+\frac{5}{9}\sin^4\theta_{\rm W},\qquad
g_R^2 = \frac{5}{9}\sin^4\theta_{\rm W}.
$$
The big ellipse in fig.\fig{NuTeV} is the NuTeV result,
obtained summing in quadrature statistical and systematic uncertainties.
$g_L$ is found to be $\sim 3\sigma$ below its SM prediction.

\medskip

The difference of the effective couplings $g^2_L-g^2_R$
(`Paschos--Wolfenstein ratio'\cite{PW}) does not depend on $r$
\begin{equation}\label{eq:PW}
R_{\rm PW} \equiv\frac{R_\nu - r R_{\bar{\nu}}}{1-r} =
\frac{\sigma(\nu {\cal N}\to \nu X)-\sigma(\bar\nu {\cal N}\to
\bar\nu X)}{\sigma(\nu {\cal N}\to \ell X) - \sigma(\bar{\nu}{\cal N}\to \bar{\ell}X)}=
 g_L^2- g_R^2 = \frac{1}{2}-\sin^2 \theta_{\rm W}.
\end{equation}
The value of $R_\nu$ measured at NuTeV is consistent with previous experiments, such as CCFR\cite{CCFR}.
Unlike CCFR, NuTeV has two separate $\nu_\mu$ and $\bar\nu_\mu$ beams:
this is the main improvement because, under the above assumptions,  allows
to get rid of the unprecisely known partonic structure of the nucleon using $R_{\rm PW}$.
The NuTeV value of $R_{\rm PW}$ is $\sim 3\sigma$ below its SM prediction.

\medskip

Doing  precision physics with iron is a delicate task.
NLO QCD corrections have not taken into account by the NuTeV collaboration.
They cancel out in the ideal $R_{\rm PW}$ observable\cite{davidson} and therefore probably
cannot explain the NuTeV anomaly
(caution is needed because what NuTeV really measures is a few $10\%$ different 
from the total cross sections which appear in the ideal PW observable).

\medskip

Parton distributions $q(x)$ are extracted from
global fits, usually performed under two simplifying assumptions:
$s(x) = \bar{s}(x)$ and $u^p(x) = d^n(x)$.
These approximation could fail, at the level of precision reached by NuTeV.
In presence of a momentum asymmetry $q^-  = \int_0^1 x [q(x) -\bar{q}(x)] dx$
the ideal PW observable shifts as
$$R_{\rm PW} = \frac{1}{2}-\sin^2\theta_{\rm W} +
 \hbox{(EW corrections)} + (1.3 + \hbox{QCD corrections}) (u^- - d^-  -  s^-).$$
Both $u^- - d^-$ (isospin violation) and $s^-$ (strange momentum asymmetry)
might produce the NuTeV anomaly compatibly with other data.

Isospin violating effects of order $u^- - d^-  \sim (m_u-m_d)/\Lambda_{\rm QCD}$
 could reconcile NuTeV with the SM compatibly with all other available data.
Some detailed computations performed replacing QCD with more
tractable phenomenological models
suggest that, due to cancellations, 
isospin-violating effects are somewhat too small\cite{u-d}.
It is not clear if a QCD computation would lead to the same conclusion.

The theoretical situation concerning $s^-$ is similar.
Since a nucleon contains 3 quarks (rather than three antiquarks)
one expects that $s$ and $\bar{s}$ carry comparable (but not equal)
fractions of the total nucleon momentum.
Non perturbative fluctuations like $p \leftrightarrow K\Lambda$
are expected to give $s$ harder than $\bar{s}$
 since $K$ is lighter than $\Lambda$.
Indeed $s^- >0$ could explain the NuTeV anomaly.
Some model computations suggest that $s^-$ is too small\cite{s-},
but again it is not clear how reliable they are.
The $s^-$ issue can also be addressed relying on experimental data.
The global parton fit in\cite{BPZ} found a hint for a
$s^- $ of the desired sign and magnitude.
However, a $s^-$ with the opposite sign was suggested by an analysis of charm production data,
obtained and analyzed by the NuTeV collaboration\cite{new}.
This analysis is unreliable for various reasons listed in a note added to\cite{davidson}.
Charm production data probe the strange content of the nucleon in a powerful way,
but we are not able of reliably guessing what they imply for $s^-$.
It seems worth performing a professional analysis, even if this not an easy task.

Finally, nuclear effects\cite{nuclear} could affect $R_\nu$ and $R_{\bar \nu}$,
but only at low momentum transfer and apparently not in
a way that allows to reconcile NuTeV data with the SM\cite{new}.
In part, nuclear effects are automatically included in the NuTeV analysis,
based on  their own parton distributions
obtained fitting only iron data.

\medskip

In conclusion, testing and eventually excluding such SM `systematic effects' which can produce the NuTeV anomaly
seems to be a difficult job\footnote{Using a different terminology,
this means that in the meantime in view of  `additional uncertainties'
NuTeV is not inconsistent with the SM.}.
Therefore, it is useful to speculate about possible new physics interpretations which might have cleaner signatures.
Unfortunately, no particularly compelling new physics with distinctive signatures has been found.
The problem is that charged  lepton couplings agree with the SM and have been
measured about 10 times more accurately than neutrino couplings.
Proposals which overcome this problem look exotic,
while more plausible possibilities work only if one
deals with constraints in a `generous' way or
introduces and fine-tunes enough free parameters.
For example, 
mixing the $Z$ boson with an extra $Z'$ boson modifies NC neutrino couplings,
but also NC couplings of charged leptons ($\ell_L$ and $\nu$ are unified in the same ${\rm SU}(2)_L$ doublet).
New physics that only affects the gauge boson propagators cannot fit the NuTeV anomaly due to the same constraints.
Neutrino oscillations do not work.
A reduction of neutrino couplings due to a $\sim 1\%$ mixing with sterile singlets 
does not work, because CC neutrino couplings have been too precisely tested
by $\mu$ decay together with precision data.
Combinations of the above effects with enough unknown parameters can work\footnote{For example,
mixing with a sterile singlet together with a heavy higgs  fits all data if one also assumes
that some extra unspecified  new physics affects $M_W$ and therefore ignores this measurement\cite{Loinaz}.
A failure of the SM fit of electroweak data
would support the case for new physics,
but at the moment we do not see any convincing problem.}.
The new physics could either be heavy with sizable couplings (so that future colliders should see it)
or light with small couplings
(e.g.\ a $Z'$ with few GeV mass and negligible mixing with the $Z$).

\section{Acknowledgements}
I learnt most of the topics discussed here thanks to collaborations with
S. Davidson, S. Forte, P. Gambino, N. Rius and F. Vissani.

\end{document}